\documentclass[12pt]{article}

\usepackage{epsfig}

\newcommand{\be}{\begin{eqnarray}}
\newcommand{\ee}{\end{eqnarray}}

\begin{document}

\thispagestyle{empty}
\title {\bf Limiting fragmentation from the Color Glass Condensate}

\author
{
Jamal Jalilian-Marian
 \\
 {\small\it Physics Department, Brookhaven National Lab,
            Upton, NY 11973, USA}\\
}

 \maketitle

\begin{abstract}

\noindent 
We show how the limiting fragmentation phenomenon can arise from
the Color Glass Condensate model of high energy QCD. We consider the
very forward rapidity region in relativistic heavy ion collisions and argue
that in this region, nucleus-nucleus collisions are similar to
proton-nucleus collisions (up to shadowing corrections). We then use
the known results for proton-nucleus cross sections to show that
it leads to the phenomenon of limiting fragmentation in the very forward
region of heavy ion collisions as observed at RHIC.

\end{abstract}

\newpage

\begin{figure}[htp]
\centering
\setlength{\epsfxsize=7.5cm}
\centerline{\epsffile{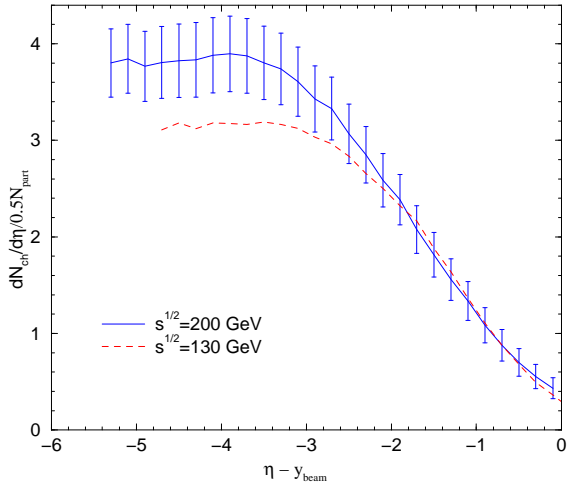}}
\caption{Limiting fragmentation observed at RHIC \cite{phobos}.}
\label{fig:limfrag}
\end{figure}

The Relativistic Heavy Ion Collider (RHIC) has opened up a
new frontier in high energy nucleus-nucleus collisions. Many
exciting and new phenomena have been observed which have
challenged theoretical models and predictions. Suppression
of high $p_t$ hadrons in mid rapidity, increase of baryon to 
pion ratio with $p_t$ and a large, constant anisotropy at high 
$p_t$ are yet to be explained satisfactorily. In the fragmentation 
region (very forward rapidity), the PHOBOS experiment \cite{phobos} 
has observed the so called limiting fragmentation phenomenon \cite{bcyy}, 
shown in Fig. (\ref{fig:limfrag}) for two different energies, which 
clearly shows most central ($0-6\%$) charged particle multiplicities are 
independent of the center of mass energy \footnote{The error bars shown 
for $\sqrt{s}=200$ GeV data are the average of positive and negative  
uncertainties published by PHOBOS \cite{phobos}.}. In this note, we show 
that the Color Glass Condensate model of high energy nuclei can lead to a 
semi-quantitative understanding of this phenomenon. 

It has been suggested that at high energies, due to high gluon
density effects, a hadron or nucleus is a Color Glass Condensate and can 
be described by semi classical methods \cite{mv}. This approach has been 
applied to heavy ion collisions at RHIC with some success \cite{dk}. Proton 
(deutron) nucleus collisions at RHIC, scheduled to begin shortly, will
greatly clarify the role and significance of the gluon saturation
at RHIC energies \cite{adfgjjm}. Here we show that limiting fragmentation
observed at RHIC can serve as yet another indication of the importance of
high gluon density effects and the Color Glass Condensate at RHIC energies.

Unlike the mid rapidity region, the fragmentation region (very forward 
rapidities) in a high energy heavy ion collision, is expected to be
quite similar to high energy proton nucleus collisions, up to shadowing
corrections\footnote{By shadowing here, we mean any modification of the
nuclear parton distributions, be it anti-shadowing, EMC effect etc.}.
This is because the Quark Gluon Plasma is expected to be formed only in 
mid rapidity and will not affect particle production in the 
forward rapidity region. Also, in the fragmentation region, one can treat the
target nucleus as a dilute system of quarks and gluons while the projectile
nucleus must be treated as a Color Glass Condensate due to its large number
of gluons. This is formally the same as a proton nucleus system treated
in \cite{adfgjjm} where one considers scattering of quarks and gluons
\cite{ajmmv} coming from the proton on the dense nucleus. For definiteness,
here we focus on quark nucleus scattering where the cross section is given by

\be
&&{d\sigma^{qA\to qX} \over d^2 q_t\,dq^-\,d^2b_t} = {1 \over (2\pi)^2}
\,\delta(q^--p^-)
\int d^2 r_t e^{i q_t \cdot r_t}
\nonumber\\ &&\!\!\!\!\times
\left[\frac{1}{N_c}
 \int d^2 R_t {\rm Tr}_c\left<
2-U(R_t +\frac{r_t}{2})
-U^\dagger(R_t -\frac{r_t}{2})
\right>_\rho
-\sigma_{\rm dipole}(r_t)\right]
\label{eq:qAcs}
\ee
with a similar equation for gluon scattering. This is the multiple 
scattering generalization of quark gluon scattering in pQCD and
is finite as $q_t \rightarrow 0$ due to higher twist effects. To relate 
this to nucleus nucleus scattering in the very forward rapidity region, 
we convolute this cross section with the quark and gluon distributions 
in the target nucleus 
\be
{d\sigma^{A\,A\to qX} \over d^2 q_t\,dq^-\,d^2b_t} = \int dx_q \,
f_{q/A}(x_q)\, {d\sigma^{qA\to qX} \over d^2 q_t\,dq^-\,d^2b_t}
\label{eq:colin}
\ee
Furthermore, since we are interested in total number of produced 
particles per unit rapidity, we will integrate over the transverse
momentum $q_t$ of the scattered quark. We emphasize the fact that 
this integral is {\it finite} and can be done {\it exactly}. It gives
\be
{d\sigma^{A\,A} \over dq^-\,d^2b_t} = 
\int dx_q \,f_{q/A}(x_q)\, {d\sigma^{qA\to qX} \over dq^-\,d^2b_t}
\label{eq:dsigdqminus}
\ee
Using $p^- = x_q \, \sqrt{s/2}$, $q^- = k^-/z$ and 
$k^-={1\over \sqrt{2}}\,m_t^h\, e^{\eta_h}$, taking advantage of 
$\delta (p^- - q^-)$ in (\ref{eq:qAcs}) to do the $x_q$ integration
and including scattering of gluons from the nucleus, we get
\be
{d\sigma^{A\,A} \over d\eta_h\,d^2b_t} \sim 
\bigg[x_q\,f_{q/A}(x_q) +x_g\,G_A(x_g)\bigg]
\label{eq:dsigdy}
\ee
where $x_q=x_g={m_t^h\over z\,m_P}e^{\eta_h - y_{beam}}$ 
and $f_{q/A}$ and $G_A$ are the quark and gluon distributions of the 
target nucleus. Experimentally, the maximum ($\eta_h - y_{beam}$) observed
at RHIC is about $+2$ units of rapidity \cite{phobos}. One can use this 
fact and that $x_q=x_g = 1$ to show that 
${m_t^h \over z\,m_P}\approx \exp(-2)$. We therefore have 
\be
x_q=x_g \approx e^{-2 \,+\, \eta_h - y_{beam}}
\label{eq:x_qg}
\ee
and all dependences on $m_t^h$ and $z$ drop out (this implicitly assumes
that hadron multiplicities are dominated by low $p_t$ pions and that $z$
does not vary much with $p_t$ for low $p_t$ pions).

\begin{figure}[htp]
\centerline{\hbox{\epsfig{figure=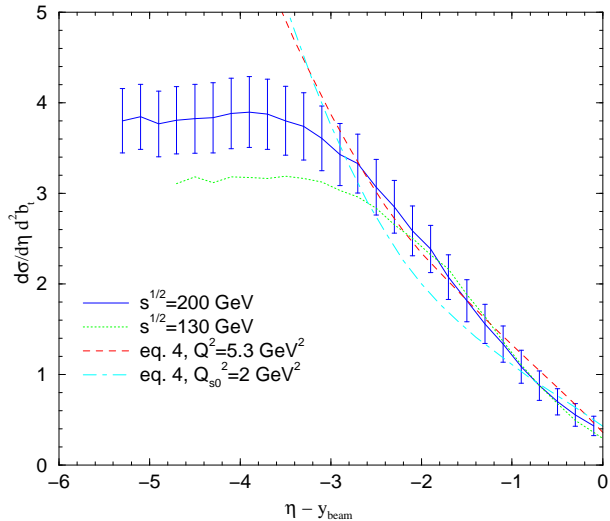,width=8cm}}}
\caption{Limiting fragmentation from (\ref{eq:dsigdy}) compared to data from
RHIC.}
\label{fig:cgclf}
\end{figure}

In Fig. (\ref{fig:cgclf}) we plot $d\sigma/d\eta_h\,d^2b_t$
from eq. (\ref{eq:dsigdy}) shifted for normalization and by the beam 
rapidity and compare with the most central ($0-6\%$) data \cite{phobos} 
from RHIC at $\sqrt{s}=200$. We have used GRV98 \cite{grv98} parton 
distributions and the EKS98 parameterization of nuclear shadowing 
\cite{eks98}.

There are a few caveats to our results. We can not predict the overall
normalizations, only the slope and have to normalize our results to the 
data at one 
reference point taken to be the target beam rapidity. It is well known 
that leading order (in $\alpha_s$) calculations of cross sections suffer 
from large scale dependence. Also, the scale dependence of nuclear parton 
distributions, and in particular the gluon distribution, is very poorly 
known due to the limited $Q^2$ coverage of fixed target experiments. The 
current parameterizations of nuclear gluon distributions are at best an 
educated guess. Unfortunately, our results are quite sensitive to the 
change of scale $Q^2$ (the scale dependence of distribution functions is
not written out explicitly in eq. \ref{eq:dsigdy}). We therefore fix 
this scale 
by requiring that eq. (\ref{eq:dsigdy}) give a reasonable fit to the RHIC 
limiting fragmentation data at $\sqrt{s}\sim 20$ GeV for a couple of units 
of rapidity. It turns out that $Q=2.3$ GeV works well. We then use the same 
scale $Q$ in (\ref{eq:dsigdy}) to predict the multiplicities at higher 
energies of $\sqrt{s}=130$ GeV and  $\sqrt{s}=200$ GeV. We also show the 
case when $Q^2=Q_s^2(y)$ as suggested \footnote{During completion of 
this note, we learned of \cite{dgs} which however focuses
on a different problem.} in \cite{dgs}. Here $y=\log 1/x$ and 
$Q_s^2(y)\equiv Q_{s0}^2\,\exp(\lambda y)$ with $Q_{s0}^2=2.0\,GeV^2$ at
mid rapidity and $\lambda=0.3$ \cite{dk}. The choice of $Q^2_{s0}=3.0\,GeV^2$
leads to a much better agreement but is disfavored \cite{dk} by RHIC data.
Alternatively and if one insists on keeping $Q_{s0}^2=2.0\,GeV^2$ at
mid rapidity, the choice of $\lambda=0.45$ leads to a good agreement
with the data but this value of $\lambda=0.45$ is too large to fit the 
HERA data (also, choosing a $x$ dependent scale in parton distributions 
would seem to violate the sum rules reflecting various conservation laws. 
Nevertheless, having this as the factorization scale might be 
theoretically tempting since otherwise there is really no hard scale left
after we integrate over all hadrons transverse momenta). We will not
pursue this further since we are not doing a detailed quantitative study
here.

There are two principle reasons why our approach should break down
as we get closer to the mid rapidity region. First, as one goes
further away from the target nucleus, high gluon density effects in the 
target nucleus become important. This will show as the growth in the
saturation scale of the target nucleus (which is of order $\Lambda_{QCD}$
right at the target nucleus rapidity). To estimate this, we use
$Q^2_s(\Delta\eta)= \Lambda^2_{QCD} \exp\,(\Delta\eta)$. As one goes
about three units of rapidity away from the target nucleus, its saturation
scale becomes appreciable ($\sim 1 $GeV) and one can not describe it as
a dilute system of partons anymore \cite{dm}. 

Another reason why this approach should break down as one gets closer to 
mid rapidity is that the classical fields of both nuclei will become
strong and the system will be very different from a proton nucleus collision. 
Also, in the mid rapidity region one will have to include 
the media effects due to the deconfined matter presumably produced in 
heavy ion collisions at RHIC. The media effects are presently not well
understood and are beyond the scope of this work.

To summarize, the underlying physics of limiting fragmentation in the 
Color Glass Condensate model is that since most particles are produced with
transverse momenta which are below the saturation scale of the
projectile nucleus (in the target nucleus reference frame), their
cross sections are transverse momentum independent (the black disk limit). 
Thus the rise of the particle multiplicities in the very forward rapidity 
region (near the target nucleus) is due to the blackness of the projectile 
nucleus and the growth of the target nucleus parton distributions with 
rapidity.

\vspace{0.2in}
\noindent {\bf Acknowledgment:}\\
The author would like to thank A.~Dumitru, L.~McLerran, R.~Pisarski 
and D.~Teaney for useful discussions. This work is supported by the 
U.S.\ Department of Energy under Contract No.\ DE-AC02-98CH10886 
and in part by a PDF from BSA.

\end{document}